# Wave vector dependent electron-phonon coupling and the charge-density-wave transition in TbTe$_3$

M. Maschek[1], S. Rosenkranz[2], R. Heid[1], A. H. Said[3], P. Giraldo-Gallo,[4,6] I.R. Fisher[5,6] and F. Weber[1]

[1] Karlsruhe Institute of Technology, Institute of Solid State Physics, 76021 Karlsruhe, Germany
[2] Materials Science Division, Argonne National Laboratory, Argonne, Illinois, 60439, USA
[3] Advanced Photon Source, Argonne National Laboratory, Argonne, Illinois, 60439, USA
[4] Geballe Laboratory for Advanced Physics and Department of Physics, Stanford University, CA 94305
[5] Geballe Laboratory for Advanced Physics and Department of Applied Physics, Stanford University, CA 94305
[6] The Stanford Institute for Materials and Energy Sciences, SLAC National Accelerator Laboratory, Menlo Park, California 94025, USA

**We present a high energy-resolution inelastic x-ray scattering investigation of the soft phonon mode in the charge-density-wave system TbTe$_3$. We analyze our data based on lattice dynamical calculations using density-functional-perturbation-theory and find clear evidence that strongly momentum dependent electron-phonon-coupling defines the periodicity of the CDW superstructure: Our experiment reveals strong phonon softening and increased phonon line widths over a large part in reciprocal space adjacent to the CDW ordering vector q$_{CDW}$ = (0, 0, 0.3). Further, q$_{CDW}$ is clearly offset from the wave vector of (weak) Fermi surface nesting q$_{FS}$ = (0, 0, 0.25) and our detailed analysis indicates that electron-phonon-coupling is responsible for this shift. Hence, we can add TbTe$_3$, which was previously considered as a canonical CDW compound following the Peierls scenario, to the list of distinct charge-density-wave materials characterized by momentum dependent electron-phonon coupling.**

The concept of a charge-density-wave (CDW) leads back to the seminal work of Peierls [1], in which he demonstrated that a one-dimensional metallic chain of atoms is unstable towards a phase transition in the presence of finite electron-phonon coupling (EPC). This phase is characterized by an oscillating electronic charge density, i.e. a CDW, and a periodic lattice distortion having the same periodicity. The corresponding ordering wave vector is of pure electronic origin and reflects the perfect nesting of the one-dimensional system at $\boldsymbol{q}_{CDW} = 2\boldsymbol{k}_F$. Since then, CDW ordered phases, also in two-dimensional materials, are commonly associated with a nesting geometry of the electronic structure. In recent years, however, it has become evident that CDW order without nesting, a possibility derived theoretically by Chan and Heine in 1973 [2], does occur in canonical CDW materials like 2H-NbSe$_2$ [3-5]. Without nesting, the periodicity of the superstructure is then defined by the momentum dependence of EPC.

The rare-earth-tritellurides ReTe$_3$ (Re = La, Ce, Gd, Tb, …) are prime examples of CDW ordered ground states. Angle-resolved photoemission spectroscopy (ARPES) measurements [6] [7] of several rare-earth tritelluride compounds including TbTe$_3$ lead to the assumption of a nesting driven CDW order and, hence, ReTe$_3$ are often used as canonical Peierls-like CDW systems. On the other hand, theoretical investigations in rare-earth tritellurides [4] revealed a nesting feature, i.e. a peak in the imaginary part of the electronic susceptibility $\chi''(\boldsymbol{q})$ at a wave vector different from the observed CDW ordering wave vector. Nonetheless, the real and imaginary parts of $\chi(q)$ in those calculations show a broad enhancement in the correct direction but the wave vector was still not correctly predicted [ $\mathbf{q}_{FS} = (0, 0, 0.25)$ compared to the observed ordering wave vector $\mathbf{q}_{CDW} = (0, 0, 0.295)$ ][8]. More importantly, as the authors pointed out, this modest enhancement was far from a divergence and as such was likely insufficient to drive the CDW instability. Recent theoretical analysis of Raman spectroscopy measurements [9] in ErTe$_3$ further corroborates that the weak nesting geometry is not identical to the ordering wave vector and enhanced electron-phonon coupling might be important.

We use high energy-resolution inelastic x-ray scattering (IXS) to investigate the lattice dynamical properties at the CDW phase transition in TbTe$_3$. We combine our experimental results with detailed lattice dynamical calculations and demonstrate that the difference between $\boldsymbol{q}_{CDW}$ and $\boldsymbol{q}_{FS}$ originates from strongly momentum dependent EPC in the case of TbTe$_3$, and likely also in other ReTe$_3$.

The experiment was performed at the XOR 30-ID (HERIX) beamline [10] [11] of the Advanced Photon Source, Argonne National Laboratory. The sample was a high-quality single crystal, grown by slow-cooling a self-flux with a CDW transition at $T_{CDW}$ = 330 K [12] and dimensions of $1 \times 1 \times 0.1 \text{mm}^3$. The components (Q$_h$, Q$_k$, Q$_l$) of the scattering vector are expressed in reciprocal lattice units (r.l.u.) (Q$_h$, Q$_k$, Q$_l$) = (h*2π/a, k*2π/b, l*2π/c) with the lattice constants $a$ = 4.316 Å, $b$ = 25.5 Å and $c$ = 4.31 Å of the orthorhombic unit cell. Phonon excitations in constant $\boldsymbol{Q}$-scans were fitted with damped harmonic oscillator functions [13]. We extracted the intrinsic phonon line width Γ of the DHO function by convoluting our fit with the experimental resolution of 1.5 meV. Thus, we obtain phonon energies $\omega_q = \sqrt{\widetilde{\omega}_q^2 - \Gamma^2}$, where $\widetilde{\omega}_q$ is the phonon energy renormalized only by the real part of the susceptibility $\chi$ and Γ is closely related to the imaginary part of $\chi$ (see discussion below).

Figure 1 shows IXS data taken well above and at the CDW transition temperature. Along the [001] direction we find two phonon branches with high intensities at $\boldsymbol{Q} = (3, 1, l)$, $0.1 \leq l \leq 0.5$: An optic branch disperses from 15 meV to

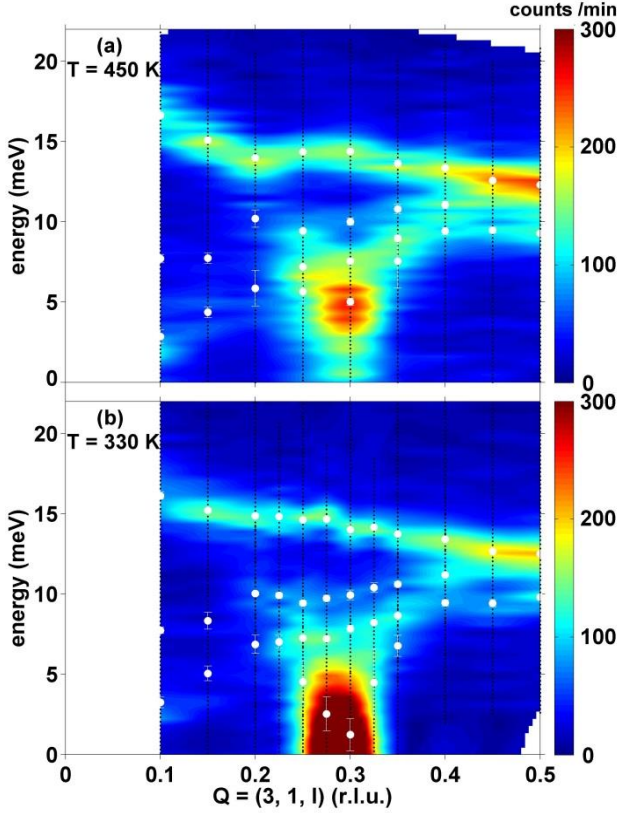

**Figure 1:** (color online) Measured intensities of transverse phonons along the [001] direction at *(a)* $T = 450$ K and *(b)* $T = T_{CDW} = 330$ K. Vertical dotted lines represent the actually performed constant-Q scans at $Q = (3, 1, l), 0.1 \leq l \leq 0.5$, summarized in color-coded contour maps. White dots are fitted phonon energies.

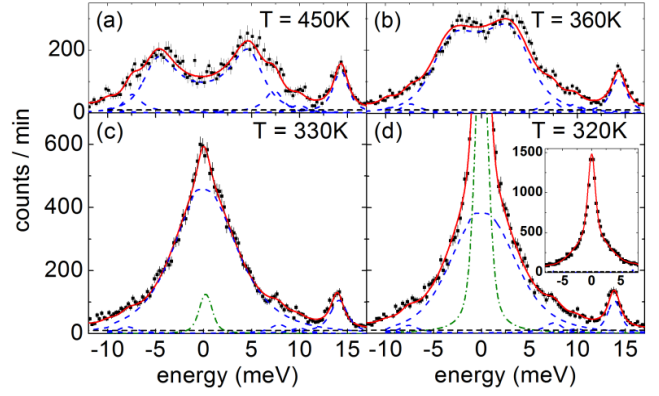

**Figure 2:** Energy scans taken at $Q = (3, 1, 0.3)$ and $320$ K $\leq T \leq 450$ K. Solid (red) lines are fits consisting of damped harmonic oscillators (inelastic, blue dashed lines) and a pseudo-Voigt function (elastic, green dashed-dotted lines). The horizontal dashed black line is the estimated experimental background. The inset in panel (d) shows the full intensity range at zero energy of the scan at T = 320 K.

12 meV in the measured wave vector range. Near half-way towards the zone boundary we observe strong scattering intensities also below 10 meV with the maximum scattering rate located at a dip in the phonon dispersion at $l = 0.3$ at $T = 450$ K [Fig. 1(a)]. The latter phonon is indeed the soft phonon mode, which is renormalized close to zero energy at $T_{CDW}$ [Fig. 1(b)], whereas the optic mode is not much affected by the temperature change. Figure 2 summarizes constant **Q** scans at $l = 0.3$, i.e. within the experimental resolution of the CDW ordering wave vector $\mathbf{q}_{CDW} = (0, 0, 0.295)$ [8], at various temperatures $320$ K $\leq T \leq 450$ K. At high temperatures [Figs. 2(a), (b)] we see that there are two additional weak phonon modes at intermediate energies, E = 7.5 meV and 10 meV. However, they are barely visible close to $T_{CDW}$ on top of the huge intensity of the soft mode and we will therefore limit our discussion to the two phonon modes dominating the spectra. The lowest energy mode at $T = 450$ K softens gradually from around 5 meV [Fig. 2(a)] to nearly zero (within the experimental error bar) at $T_{CDW} = 330$ K [Fig. 2(c)]. Here, the elastic scattering of the CDW superstructure peak emerges and dominates the spectrum at 320 K [Fig. 2(d)].

Figure 3 summarizes the measured dispersion at $T_{CDW}$ [Fig. 3(a)], the observed temperature dependence of the energies [Fig. 3(b)] and the line widths Γ [Fig. 3(c)] of the soft mode and the optic mode at $\mathbf{q}_{CDW}$. The soft phonon mode dispersion is clearly visible with a dip around $\mathbf{q}_{CDW}$ and the extracted temperature dependences show the expected softening [Fig. 3(b)] and broadening [Fig. 3(c)] strongest at $T_{CDW}$. For comparison, we show the corresponding practically temperature independent values for the optic mode at $\mathbf{q}_{CDW}$.

For a more detailed analysis we performed *ab-initio* calculations for the lattice dynamical properties based on *density-functional-perturbation-theory* (DFPT) using the high-temperature orthorhombic structure present at $T > T_{CDW}$. In TbTe$_3$, the $f$ states of Tb are localized and shifted away from the Fermi energy. Therefore, they are expected to play no role for the physics of the CDW [7] [14]. To avoid complications with the $f$ states in standard density-functional theory, we have performed our calculations for LaTe$_3$ in the framework of the mixed basis pseudopotential method [15]. Norm-conserving pseudopotentials were constructed following the scheme of Vanderbilt including *5s* and *5p* semicore states of La in the valence space but excluding explicitly *4f* states [16]. The basis set consisted of plane waves up to 20 Ry complemented with local functions of s, p, and d symmetry at the La sites. For the exchange-correlation functional the local-density approximation (LDA) was applied [17]. DFPT as implemented in the mixed basis pseudopotential method [18] was used to calculate phonon frequencies and electron-phonon coupling. An orthorhombic 24 × 8 × 24 **k**-point mesh was employed in the phonon calculation, whereas an even denser 48 × 12 × 48 mesh was used in the calculation of phonon line widths to ensure proper convergence. This was combined with a standard smearing technique using a Gaussian broadening of 0.24 eV.

The calculated dispersion is in good agreement with our data at $T_{CDW}$ [Fig. 3(a)], i.e. DFPT predicts the structural instability at the correct wave vector position. Hence, we confirm that *4f* states are not relevant for CDW formation in TbTe$_3$ in agreement with previous reports [4,7]. The calculated intensities for the phonon modes at $\mathbf{Q} = (3, 1, l)$, $0.1 \leq l \leq 0.5$, are in excellent agreement with our IXS data demonstrating that the predicted transverse optic mode is indeed the CDW soft mode. Hence, we can use DFPT to follow the character of the soft mode along the [001] wave vector direction. We find that the soft mode at $\mathbf{q}_{CDW}$ in fact has an optic character. It starts out at the zone center at 15 meV (with zero spectral weight in the measured Brillouin zone) and slowly bends downward and exhibits two anti-crossings with lower energy branches of the same symmetry in short order just above and below an energy of $E$ = 5 meV. Beyond $\mathbf{q}_{CDW}$ it disperses upwards again. In the following we refer to the described dispersion as the soft mode dispersion [marked red in Fig. 3(a)]. The dispersion of the high-energy optic branch is marked in blue and shows also good agreement with experiment.

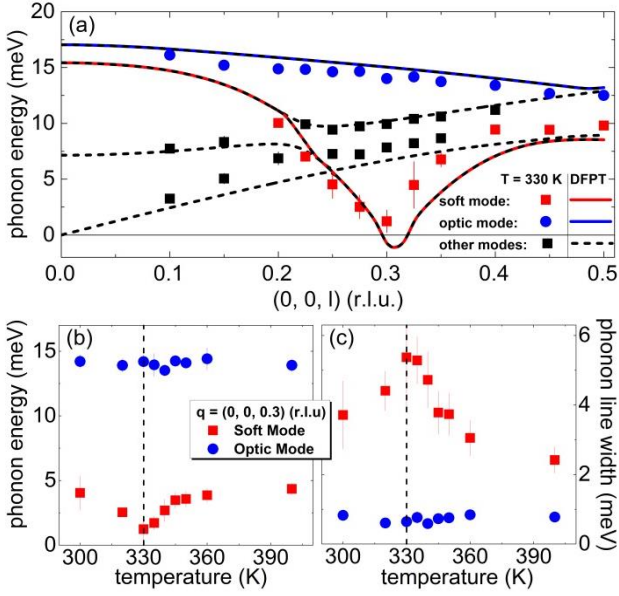

**Figure 3:** *(a)* Comparison between calculated (dashed lines) and observed (*T* = 330 K) energies of transverse phonon modes along the [001] direction (squares). Red and blue lines mark the dispersion of the soft phonon mode character and a strong optic mode, respectively. The temperature dependences of the phonon energy and line width of the marked modes at $q = (0, 0, 0.3)$ are shown in panels *(b)* and *(c)*, respectively. The dashed line denotes the transition temperature $T_{CDW}$ = 330 K.

In the standard theory of CDW order outlined by Chan and Heine [2] the properties of the soft phonon mode are directly connected to the electronic structure via the real and imaginary parts of the electronic susceptibility, $\chi'(\mathbf{q})$ and $\chi''(\mathbf{q})$. Whereas $\chi'(\mathbf{q})$ is mostly reflected in the renormalization of the phonon energy, $\chi''(\mathbf{q})$ is in first order proportional to the line width of the soft mode, which reflects the reduced life time due to EPC. Striking is the difference between the minimum in the calculated soft mode dispersion [Fig. 3(a), indicated by $\mathbf{q}_{SM}$ in Fig. 4(a)] and the maxima of both $\chi'(\mathbf{q})$ and $\chi''(\mathbf{q})$ [Fig. 4(a), $\chi'(\mathbf{q})$ and $\chi''(\mathbf{q})$ reproduced from Ref. [4]], which are clearly separated. According to Ref. [2] the strength of the phonon renormalization is proportional to the product $\chi(\mathbf{q}) \cdot \eta_\mathbf{q}^2$, which includes the squared averaged EPC matrix element $\eta_\mathbf{q}^2$. Thus, we assign the observed difference to a strongly momentum dependent $\eta_\mathbf{q}^2$.

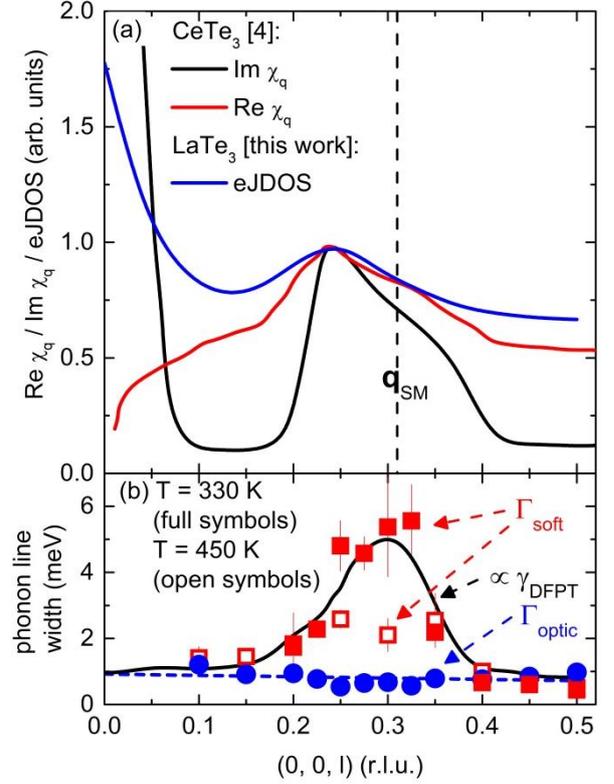

**Figure 4:** *(a)* Comparison of calculated real and imaginary part of the non-interacting electronic susceptibility $\chi_q$ for CeTe$_3$ extracted from Ref. [8] with the electronic joint density-of-states from DFPT in this work. The vertical dashed line indicates the position of the soft mode in the DFPT calculation. *(b)* Wave vector dependences of line widths of the soft mode $\Gamma_{soft}$ (red squares) and optical mode $\Gamma_{optic}$ (blue dots) at $T_{CDW}$ = 330 K. Open symbols denote the values of $\Gamma_{soft}$ at $T = 450$ K. The solid line represents a DFPT calculation of the line width of the soft mode. For comparison it has an offset (dashed blue line) and is scaled (see text).

From our own calculations, we can extract the electronic joint density-of-states (eJDOS), which is directly related to $\chi''(\mathbf{q})$ and reflects the number of electronic states at the Fermi surface connected by a particular wave vector, i.e., the tendency to Fermi surface nesting. We find good agreement of the wave vector position of the maxima of the eJDOS and $\chi''(\mathbf{q})$ from Ref. [4] [Fig. 4(a)]. Based on this, we can calculate the electronic contribution to the phonon line width $\gamma$ [Fig. 4(b)]. We again find an offset, now between the wave vector positions of the maximum of $\gamma$ and the eJDOS, where the former agrees well with the minimum of the soft mode dispersion [indicated by $\mathbf{q}_{SM}$ in Fig. 4(a)]. Hence, the momentum dependence of $\gamma$ is set by $\eta_\mathbf{q}^2$ and the number of available electronic decay channels, i.e. the eJDOS [19]. Thus, the presence of a strongly momentum dependent EPC matrix element $\eta_q^2$ is demonstrated within our *ab-initio* calculation.

Experimentally, the wave vector range with large phonon line widths Γ agrees well with the prediction of DFPT [Fig. 4(b)] corroborating our interpretation. Quantitatively, DFPT underestimates Γ by a factor of about 20. Indeed, we found in several studies that DFPT underestimates the experimental line widths for soft phonon modes in the CDW compounds NbSe$_2$ [5,20] and TiSe$_2$ [21] but also in the superconductor YNi$_2$B$_2$C [22]. However, DFPT predictions were typically a factor of only two below the observed values in these compounds, and were assigned to anharmonic contributions disregarded in DFPT. The much larger discrepancy for TbTe$_3$ originates in a much smaller calculated $\gamma$, whereas the experimental values of Γ in TbTe$_3$ and NbSe$_2$ [5] and also TiSe$_2$ [21] are comparable. Although it seems that the CDW in the rare-earth tritellurides is mainly related to Te states at the Fermi level we cannot exclude that inclusion of *4f* states into our calculations is required for an improved quantitative description. This cannot be tackled easily and is subject of future research.

Here, we argue that the obvious mismatch between maxima in $\chi(q)$ on the one side and the wave vector position of the soft mode and the largest line width at $q_{CDW}$ on the other side, in particular within the DFPT theory itself, demonstrate the defining influence of momentum dependent EPC in TbTe$_3$. A similar proposal has been made based on Raman measurements in ErTe$_3$ [9]. Hence, a *q* dependent EPC is likely relevant for the whole family of rare-earth tritellurides, which is also supported by the fact that we can model the behavior of TbTe$_3$ rather well without taking into account *4f* electrons. So far, the importance of momentum dependent EPC was shown for compounds, for which the nesting scenario was questioned already for a long time [5,23]. Our results for TbTe$_3$ demonstrate that EPC needs to be investigated even if the periodicity of the CDW order is "roughly" explained by the electronic structure.

In conclusion we reported inelastic x-ray measurements and DFPT calculations of the soft phonon mode on entering the CDW ordered phase in TbTe$_3$. Our observation of a broad softening and enhanced phonon line widths over a large range of wave vectors is well explained by DFPT lattice dynamical calculations. A more detailed analysis shows that the wave vector dependency of EPC in TbTe$_3$ is defining the ordering wave vector **q**$_{CDW}$ in contrast to the standard Fermi surface nesting scenario.

M.M. and F.W. were supported by the Helmholtz Society under contract VH-NG-840. Work at Argonne was supported by U.S. Department of Energy, Office of Science, Office of Basic Energy Sciences, under Contract No. DE-AC02-06CH11357. The work at Stanford University was supported by the DOE, Office of Basic Energy Science, under Contract No. DE-AC02-76SF00515.